\documentclass[10pt]{iopart}
\usepackage[a4paper, total={6in, 9in}]{geometry}
\usepackage[utf8]{inputenc}
\usepackage[T1]{fontenc}
\usepackage{comment}        
\usepackage{subcaption}     
\usepackage{graphicx}       
\usepackage{relsize}        
\usepackage{float}          
\usepackage{amsmath}        
\usepackage{amssymb}        
\usepackage{enumitem}       
\usepackage[style=numeric, sorting=none]{biblatex} 
\begin{document}
\title{Electron-positron pair production in the collision of real photon beams with wide energy distributions}

\author{L. Esnault$^1$, E. d'Humières$^1$, A. Arefiev$^2$, and X. Ribeyre$^1$}

\address{$^1$Univ. Bordeaux-CNRS-CEA, Centre Lasers Intenses et Applications, UMR 5107, 33405 Talence, France}
\address{$^2$Department of Mechanical and Aerospace Engineering, University of California at San Diego, La Jolla, California 92093, USA}
\eads{\mailto{leo.esnault@u-bordeaux.fr}, \mailto{xavier.ribeyre@u-bordeaux.fr}}

\begin{abstract}
The creation of an electron-positron pair in the collision of two real photons, namely the linear Breit-Wheeler process, has never been detected directly in the laboratory since its prediction in 1934 despite its fundamental importance in quantum electrodynamics and astrophysics.
In the last few years, several experimental setup have been proposed to observe this process in the laboratory, relying either on thermal radiation, Bremsstrahlung, linear or multiphoton inverse Compton scattering photons sources created by lasers or by the mean of a lepton collider coupled with lasers.
In these propositions, the influence of the photons' energy distribution on the total number of produced pairs has been taken into account with an analytical model only for two of these cases.
We hereafter develop a general and original, semi-analytical model to estimate the influence of the photons energy distribution on the total number of pairs produced by the collision of two such photon beams, and give optimum energy parameters for some of the proposed experimental configurations.
Our results shows that the production of optimum Bremsstrahlung and linear inverse Compton sources are, only from energy distribution considerations, already reachable in today's facilities. 
Despite its less interesting energy distribution features for the LBW pair production, the photon sources generated via multiphoton inverse Compton scattering by the propagation of a laser in a micro-channel can also be interesting, thank to the high collision luminosity that could eventually be reached by such configurations.
These results then gives important insights for the design of experiments intended to detect linear Breit-Wheeler produced positrons in the laboratory for the first time.
\end{abstract}

\submitto{\NJP}
\maketitle
\section{Introduction}

\begin{figure}[h]
    \centering
    \includegraphics[width=\linewidth]{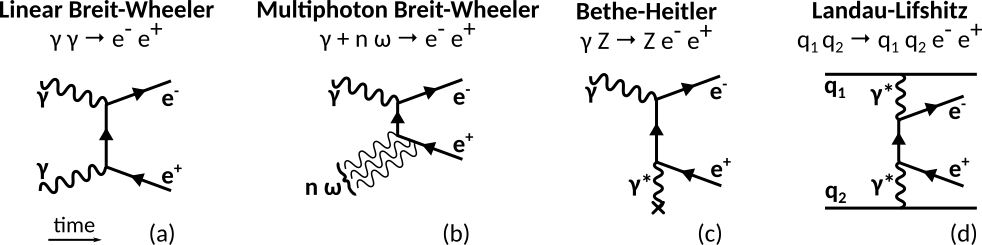}
    \caption{Examples of electron-positron pair production by photon-photon collision processes. $\gamma$ stands for high energy real photon, $\omega$ for low energy real photon and $\gamma^{*}$ for virtual photon.}
    \label{fig:1-QED_LBW}
\end{figure}

Electron-positron pair production by the collision of two real photons (linear Breit-Wheeler process, LBW, Figure \ref{fig:1-QED_LBW}a) is a pure quanta effect, and is believed to play an important role in the high energy photon opacity of the Universe \cite{nikishov_1961} and in extreme astrophysical events \cite{ruffini_2010}, such as in the neighbourhood of active galactic nuclei \cite{svensson_1987} or pulsars polar caps \cite{zhang_1998}.
Despite being one of the most basic processes of quantum electrodynamics (QED), it has however never been detected directly in the laboratory since its prediction in 1934 \cite{breit_1934}, mainly because of the absence of sufficiently high flux real $\gamma$ ray sources. 

Indirect measurements of this process have however been performed in the laboratory through the measurement of several close mechanisms, such as the $e^-e^+$ annihilation into two photons \cite{dirac_1930, klemperer_1934} (inverse process), or the pair production in the collision of one high energy photon with several low energy photons \cite{reiss_1962, burke_1997} (multiphoton Breit-Wheeler process, Figure \ref{fig:1-QED_LBW}b), the collision of a real photon with an electric field, interpreted in the theoretical framework of QED as a virtual photon \cite{bethe_1934, anderson_1933} (such as the Bethe-Heitler process when the electric field arises from atomic nuclei, Figure \ref{fig:1-QED_LBW}c), or through the collision of two virtual photons in charged particle collisions \cite{landau_1934,balakin_1971} (Landau-Lifshitz process for charges $q_1$ and $q_2$ in Figure \ref{fig:1-QED_LBW}d).
These virtual photon collision processes have already been widely investigated \cite{budnev_1975, hubbell_2006} but are still of great interest nowadays \cite{klein_2020,starcollaboration_2019}.

From a theoretical perspective, these processes are linked to real photon pair production because under some particular conditions constituting the equivalent photon approximation \cite{budnev_1975, kessler_1974} (also called Williams-Weiszacker approximation), virtual photons can be treated as real ones, allowing to use the LBW cross section to compute the virtual photon pair production probability \cite{greiner_2009}.

Real photon collision is however also a topic of interest \cite{chou_2018, takahashi_2019}.
Especially, since 2014, propositions have emerged to study the LBW process in the laboratory with controlled photon sources for the first time. 
These propositions relies either on the collision of two identical Bremsstrahlung sources produced by laser-solid interaction \cite{ribeyre_2016}, two identical multi-photon inverse Compton sources produced by laser interacting either with an homogenenous target \cite{ribeyre_2016} or propagating in a micro-channel of sub-solid density \cite{yu_2019, jansen_2018, wang_2020, wang_2020b, he_2020}, by the collision of two identical linear inverse Compton sources produced by the interaction of lasers with electron beams in a lepton collider \cite{drebot_2017}, or by the interaction of a laser-produced Bremsstrahlung source with a thermal X-ray bath \cite{pike_2014} or with an X-ray laser \cite{golub_2020}.

In these laboratory or astrophysical situations, the real $\gamma$ ray sources can have a more or less wide energy distribution (e.g. black-body like, Bremsstrahlung, ...).
To estimate the total number of LBW-produced electrons or positrons, one should then sum up the contribution of all the possible $\gamma$ binary collisions (i.e. the collision between only two photons of determined energies). 
Moreover, the optimization of these features are of particular importance in the design of LBW laboratory experiments.
The computation of pair number by the mean of simple analytical models taking into account the incoming photons energy distributions is however not straightforward in the general case, and these kinds of analytical or semi-analytical estimates are yet available only for the pair number in two laboratory propositions \cite{pike_2014, golub_2020}. Other estimates rely either on single energy approximation \cite{ribeyre_2016, he_2020} for the incoming photon sources, or on numerical computations \cite{yu_2019, wang_2020, drebot_2017}.

Other situations have however been extensively studied in the astrophysics community, such as the interaction of a high energy photon with an isotropic, low energy photon gas with black-body \cite{nikishov_1961, zhang_1998, gould_1967, voisin_2018}, power-law \cite{gould_1967, bonometto_1971}, Wien-type \cite{schlickeiser_2012} or synchrotron \cite{voisin_2018} energy distribution, or in the collision of two isotropic photon sources with power-law \cite{svensson_1987, bottcher_1997} energy distributions. 
In these analytical or semi-analytical models, at least one of the two photon sources is supposed to be isotropic or quasi-isotropic, and a large asymmetry in the incoming photons energies is often assumed to run the calculations, implying that these models can not be directly applied to the collision of two directional photon beams having both a similar and wide energy distribution.

The aim of this paper is to give a general, semi-analytical method to estimate the total number of positrons (or equivalently electrons) produced by the LBW process in the collision of two photon beams, given both of the beams have simple energy distributions but no angular divergence (i.e. both of them are perfectly directional sources). These estimates can be helpful tools to compare various input $\gamma$ sources and to determine the most suitable ones to observe the LBW process in the laboratory for the first time.


\section{Energy distribution effects on pair number}
To estimate the total production of electron-positron pairs in the interaction of two $\gamma$ beams, we need to consider all the possible $\gamma$ binary collisions. After presenting a general semi-analytical method, we will apply our model to three laboratory-relevant configurations, and give a numerical fit of the obtained data.
\subsection{Principle}

\begin{figure}[t]
    \begin{minipage}[b]{0.4\linewidth}
        \centering
        \includegraphics[width=\textwidth]{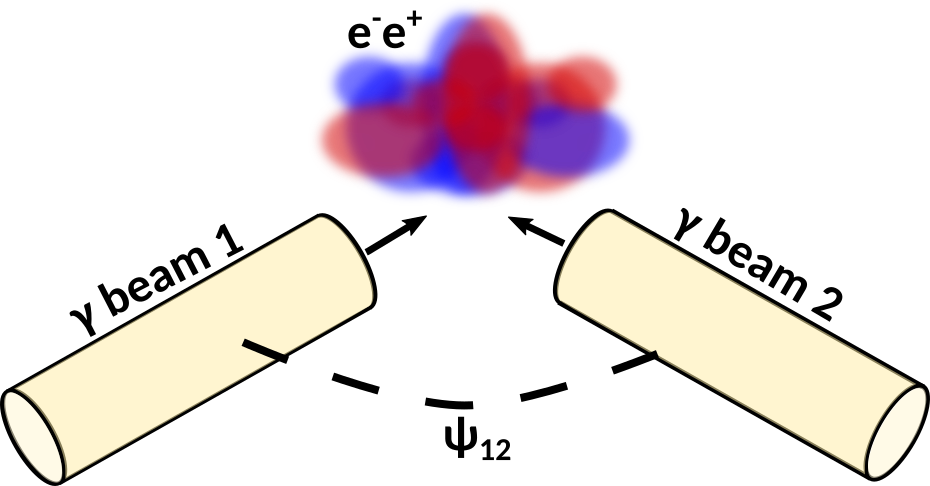}
        \caption{Principle of beam-beam collision. $\psi_{12}$ is the collision angle between the two photon beams denoted by the yellow cylinders. The produced $e^-e^+$ pairs are represented by the blue and red cloud.}
        \label{fig:2.1-beam-beam_collision}
    \end{minipage}
    \hfill
    \begin{minipage}[b]{0.5\linewidth}
        \centering
        \includegraphics[width=\textwidth]{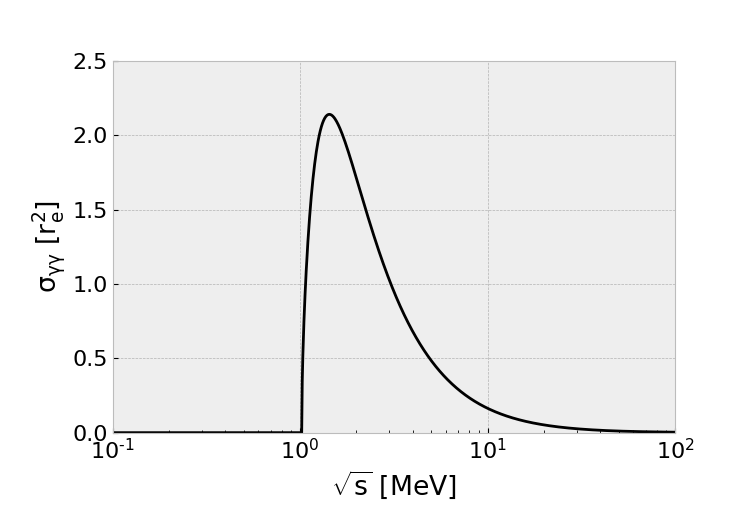}
        \caption{Linear Breit-Wheeler cross section (in $r_e^2$), defined in equation (\ref{eq:2.1-LBW_cross_section}), plotted against $\sqrt{s}=E_{CM}$ (in MeV).}
        \label{fig:2.1-LBW_cross_section}
    \end{minipage}
\end{figure}

In this paper, we consider the collision of two photon beams with no angular divergence, i.e. that all the photons in a given beam moves parrallel to each other, such as illustrated in figure \ref{fig:2.1-beam-beam_collision}. In this case, one can compute the LBW volumic production rate of positrons from \cite{gould_1971,landau_1975}: 
\begin{equation}
    \dfrac{d n_+}{dt} = \iint c ~ (1 - \cos \psi_{12}) ~ \dfrac{d n_1}{d E_1} ~ \dfrac{d n_2}{d E_2} ~ \sigma_{\gamma\gamma} ~ d E_1 ~ d E_2 \rm ,
    \label{eq:2.1-production_rate}
\end{equation}
with $n_+$ the number density of positrons, $n_i$ the number densities of photons for beam $i$  ($i=\{1,2\}$), $dt$ an infinitesimal timestep, $E_i$ the energies of the photons of beam $i$, $c$ the light velocity in vacuum, $\psi_{12}$ the angle between the photons (which is the same in all the photons binary collisions), and $\sigma_{\gamma\gamma}$ the LBW cross section given by \cite{greiner_2009}:
\begin{equation}
    \sigma_{\gamma\gamma}(s)= 4 \pi r_e^2 \frac{m_e^2 c^4}{s} \left[ \left(2 + \frac{8 m_e^2 c^4}{s} - \frac{16 m_e^4 c^8}{s^2}\right) \ln \frac{\sqrt{s} + \sqrt{s - 4 m_e^2 c^4}}{2 m_e c^2} - \sqrt{1 - \frac{4 m_e^2 c^4}{s}}\left(1 + 4 \frac{m_e^2 c^4}{s}\right)\right] \rm ,
    \label{eq:2.1-LBW_cross_section}
\end{equation}
where $r_e \approx 2.8 \times 10^{-15}$ m is the classical electron radius, $m_e$ the electron mass, and $s=2 E_1 E_2 (1 - \cos \psi_{12})=E_{CM}^2$ the squared invariant mass Mandelstam variable, with $E_{CM}$ being the center of momentum energy. 
This cross section, plotted against $\sqrt{s}=E_{CM}$ in figure \ref{fig:2.1-LBW_cross_section}, depends on the individual energies of the photons involved in each binary collision, has a pair production threshold given by the condition $\sqrt{s} \geq 2 m_e c^2$, and a maximum for $\sqrt{s} \approx 1.44$ MeV.

We consider here that the $\gamma$ energy distribution in a given photon beam is constant and homogeneous, i.e. that we can express $d n_i/d E_i$ in term of a product of a spatio-temporal part and an energy distribution function (EDF):
\begin{equation}
    \dfrac{dn_i(\textbf{x}, t, E_i)}{dE_i} = N_i \times \rho_i (\textbf{x}, t) \times f_i (E_i) ~ \rm ,
    \label{eq:2.1-homogeneous_beam}
\end{equation}
with $\textbf{x}$ the position vector, $t$ the time, $N_i$ the total number of photons in beam $i$, $\rho_i$ their normalized density and $f_i$ their energy distribution function. We can then write the total number of positrons produced in a single beam-beam collision $N_+$:
\begin{equation}
    N_+ = \iiiint \dfrac{d n_+}{dt} ~ d^3 V ~ dt = \mathcal{L}_{12} \times \sigma_{\gamma\gamma}^{int} \rm ,
    \label{eq:2.1-N+_definition}
\end{equation}
in term of the single collision geometric luminosity \cite{gould_1971,herr_2006} of the beam-beam collision $\mathcal{L}_{12}$:
\begin{equation}
    \mathcal{L}_{12} = c ~ (1 - \cos \psi_{12}) ~ N_1 ~ N_2 \iiiint_{-\infty}^{+\infty} \rho_1 ~ \rho_2 ~ d^3 V ~ dt \rm ,
\end{equation}
where $d^3 V$ is an infinitesimal volume, and an integrated cross section $\sigma_{\gamma\gamma}^{int}$ taking into account the energy coupling of the beams for given $f_1$ and $f_2$:
\begin{equation}
    \sigma_{\gamma\gamma}^{int} = \iint_{0}^{+\infty} f_1 ~ f_2 ~ \sigma_{\gamma\gamma} ~ dE_1 ~ dE_2 \rm .
    \label{eq:2.1-ECS_definition}
\end{equation}

Thus, each of these two quantities can gives interesting insights for the maximization of the number of LBW pairs produced in the interaction of such photon beams. 

The geometric luminosity $\mathcal{L}_{12}$ depends only on the total number of photons and on geometric factors (spatio-temporal profiles of the beams, collision angle, ...). This quantity is then highly dependant on the $\gamma$ production mechanism, and on the choice of collision angle and collision distance. Its computation by the mean of analytical model is in general not straightforward (especially for large collision angle), and will not be developped in this specific paper (see e.g. reference \cite{herr_2006} for more informations). For the rough estimates that will be made at the end of this paper, we will compute the luminosity as being the product of the number of photons in each beam divided by the typical surface of the photon beams at the collision point.

On the other hand, the integrated cross section depends only on energetic factors, and can then be computed once and for all for given photon beams energy distributions $f_1$ and $f_2$. 
This quantity, defined the same way as cross sections involving virtual photons under the equivalent photon method \cite{budnev_1975, kessler_1974}, is a scalar taking into account all the $\gamma$ energy couplings ponderated by the pair production probability (modelised by the two photons LBW cross section $\sigma_{\gamma \gamma}$). The computation of $\sigma_{\gamma\gamma}^{int}$ for various input photon EDF can then be of great help to compare their relative efficiency for the application of LBW pair production. 

Depending on the form of $f_1$ and $f_2$, the calculation of $\sigma_{\gamma\gamma}^{int}$ has in general no simple solution. Hereafter we will restrict ourselves to EDF that can be written in the form:
\begin{equation}
    f_i(E_i, K_i) = \dfrac{g_i(E_i/K_i)}{K_i} \rm ,
    \label{eq:2.1-g_definition}
\end{equation}
and therefore, the EDF normalization condition writes:
\begin{equation}
    \int_0^{+\infty} f_i (E_i, K_i) ~ dE_i = \int_0^{+\infty} g_i (E_i/K_i) ~ d(E_i/K_i) = 1 \rm ,
    \label{eq:2.1-f_normalization}
\end{equation}
with $K_i$ being a characteristic energy for beam $i$ and $g_i$ any one variable distribution satisfying the normalization condition given by equation (\ref{eq:2.1-f_normalization}). Some examples can be found in table \ref{tab:2.1-ED_definitions}.

We can then rewrite equation (\ref{eq:2.1-ECS_definition}) with proper dependencies:
\begin{equation}
    \sigma_{\gamma\gamma}^{int}(K_1, K_2, \psi_{12}) = \iint_0^{+\infty} \dfrac{g_1(E_1/K_1)}{K_1} \dfrac{g_2(E_2/K_2)}{K_2} ~ \sigma_{\gamma\gamma}(E_1, E_2, \psi_{12}) dE_1 dE_2 ~ .
    \label{eq:2.1-ECS_with_dependencies}
\end{equation}

To simplify this expression, we will express the LBW cross section defined by equation (\ref{eq:2.1-LBW_cross_section}) only in term of the parameter $s=2 E_1 E_2 (1 - \cos \psi_{12})$. If we define new variables $\eta = 2 K_1 E_2 (1 - \cos \psi_{12})$ and $\zeta= 2 K_1 K_2 (1 - \cos \psi_{12})$, we can notice that $E_1/K_1 = s/\eta$ and $E_2/K_2 = \eta/\zeta$. We can then rewrite the integrand of equation (\ref{eq:2.1-ECS_with_dependencies}) only in term of variables $s$ and $\eta$ ($\zeta$ being here a constant parameter), and the Jacobian determinant for this substitution writes:
\begin{equation}
    J = 
    \begin{vmatrix}
    \dfrac{\partial E_1}{\partial s}    & \dfrac{\partial E_1}{\partial \eta} \\
    \vspace{-2mm}\\
    \dfrac{\partial E_2}{\partial s}    & \dfrac{\partial E_2}{\partial \eta} \\
    \end{vmatrix}
    =
    \begin{vmatrix}
    \dfrac{K_1}{\eta}       & -K_1\dfrac{s}{\eta^2} \\
    \vspace{-2mm}\\
    0                       & ~ \dfrac{K_2}{\zeta} \\
    \end{vmatrix}
    = \dfrac{K_1}{\eta} \dfrac{K_2}{\zeta} \rm .
\end{equation}

Using the change of variable theorem, $\sigma_{\gamma\gamma}^{int}$ can then be written:
\begin{equation}
    \sigma_{\gamma\gamma}^{int}(\zeta) = \dfrac{1}{\zeta} \int_0^{+\infty} g_2(\eta/\zeta) \left[\dfrac{1}{\eta} \int_0^{+\infty} g_1(s/\eta) \sigma_{\gamma\gamma}(s) ds \right] d\eta ~ \rm .
    \label{eq:2.1-ECS_g_g}
\end{equation}

Within this context, the integrated cross section thus depends only on the single parameter $\zeta=2 K_1 K_2 (1 - \cos \psi_{12})$, where $\sqrt{\zeta}$ can be interpreted as a characteristic energy of the beam-beam collision. From energy distribution considerations, the value of $\zeta$ itself can then completely characterize the interaction of such beams, and the maximization of $\sigma_{\gamma\gamma}^{int}$ could be done by properly choosing the value this parameter. The value of $\zeta$ that maximizes $\sigma_{\gamma\gamma}^{int}$ will then be noted $\zeta'$ in the following, and can be achieved for any $K_1$, $K_2$ and $\psi_{12}$ providing that the relation:
\begin{equation}
    2 K_1 K_2 (1 - \cos{\psi_{12}}) = \zeta' 
    \label{eq:2.1-optimum_S}
\end{equation}
is satisfied. This relashionship can then be of great help to determine the most suitable characteristic energies and collision angle for the maximization of the produced number of pairs, as it will be shown with some examples later in this paper.

For the collision of two real photon beams with arbitrary energy distributions and no angular divergence, we have seen that, in some conditions, we can express the total number of pairs as the product of a geometric luminosity and a so-called integrated cross section. This integrated cross section takes into account all the possible $\gamma$ energy couplings, and can be calculated once and for all for given photon energy distributions. For simple photon energy distributions, it can be reduced to a one-variable function, allowing to optimise the photons energy distribution characteristics and collision angle easily.

We will now apply equation (\ref{eq:2.1-ECS_g_g}) in the context of two directional photon beams collision for LBW experiments. We recall that the $\gamma$ energy distribution function of each beam $i=\{1,2\}$, $f_i$, must be written in the form of equation (\ref{eq:2.1-g_definition}) and should be normalizable to unity, such as stated in equation (\ref{eq:2.1-f_normalization}).

\subsection{Application to beam-beam laboratory collisions}

In this section, we will use the definition of the integrated cross section given by equation (\ref{eq:2.1-ECS_g_g}) to model some of the experimental propositions previously described. Especially, we will focus on the interaction of two laser-based Bremsstrahlung \cite{ribeyre_2016} photon sources, two multiphoton inverse Compton scattering $\gamma$ sources produced by a laser propagating in a micro-channel \cite{wang_2020}, and two inverse-Comptonized laser photon beams in a lepton collider \cite{drebot_2017}. After defining the form of the $g_i$ functions to model these propositions, we will discuss the effects of such energy distributions on the total number of produced pairs for these experiments. A fitting function is then given to reproduce the numerically computed data. The model could also be adapted to other experimental or astrophysical situations, provided that the photon beams are supposed to be perfectly directional, and that the energy distribution functions satisfies the requirements of equations (\ref{eq:2.1-g_definition}) and (\ref{eq:2.1-f_normalization}).

\subsubsection{Energy distribution functions for selected photon sources}
In this paper, we will then illustrate our model by considering especially Bremsstrahlung photon sources, multi-photon inverse Compton photon sources and linear inverse Compton sources. Given some simplifications and approximations, the energy distribution of such $\gamma$ sources can be approximated by simple functions satisfying simultaneously the requirements of equations (\ref{eq:2.1-g_definition}) and (\ref{eq:2.1-f_normalization}):

\begin{enumerate}[label=\roman*. ]
    \item To model the energy distribution of Bremsstrahlung $\gamma$ sources, we will use exponential functions, such as defined in table \ref{tab:2.1-ED_definitions}. This kind of function, satisfying both equations (\ref{eq:2.1-g_definition}) and (\ref{eq:2.1-f_normalization}), is indeed widely used in the litterature to characterize the high energy tail of such $\gamma$ source \cite{norreys_1999, zulick_2013, henderson_2014}, with a slope that will be called "characteristic energy" here (sometimes called "effective temperature" in the litterature). Analytical formulae also exists to describe the $\gamma$ energy distributions produced by Bremsstrahlung on thin \cite{shkolnikov_1997} or thick \cite{matthews_1973, findlay_1989, tsai_1974} targets, but they usually do not satisfy either the form of equation (\ref{eq:2.1-g_definition}) or (\ref{eq:2.1-f_normalization}), or none of the two. We will also briefly discuss the use of a sum of exponentials at the end of this section. 

    \item The energy distribution of the photons produced via the multi-photon inverse Compton scattering process by a laser propagating in a micro-channel is modeled here by the product of a power-law with an exponential (see the definition in table \ref{tab:2.1-ED_definitions}). This kind of function, satisfying both equations (\ref{eq:2.1-g_definition}) and (\ref{eq:2.1-f_normalization}), can reproduce reasonably well the low energy part of the data obtained numerically for the 2 PW and 4 PW cases of the reference \cite{wang_2020}, and is especially accurate from 10 keV to 10 MeV range (data obtained by courtesy of T Wang). Using the sum of two of such functions, the simulated data can be well reproduced from 10 keV to about 100 MeV.
    This kind of photon sources are sometimes called "synchrotron-like" in the laser-plasma community because of the similarities between these kind of sources and the synchrotron emission of radiation \cite{lau_2003}. However, the formulae describing the energy distribution of the photon emission by an electron in instantaneous circular motion \cite{jackson_2009} is less accurate to reproduce the data than the function given in table \ref{tab:2.1-ED_definitions}.

    \item The photon sources produced via the linear inverse Compton process by the scattering of an electron beam with a laser are modeled by a step function, which is constant up to a cutoff energy (see the definition in table \ref{tab:2.1-ED_definitions}). This rough modeling can approximately reproduce the data obtained numerically in the reference \cite{drebot_2017} (see their figure 4).
    More precise analytical formulae to describe such photon source are also available in the litterature \cite{fargion_1997}, but they usually do not satisfy the requiremenents of equations (\ref{eq:2.1-g_definition}) and (\ref{eq:2.1-f_normalization}).
    
\end{enumerate}

\begin{figure}[t]
    \begin{minipage}[h]{0.45\linewidth}
        \centering
        \includegraphics[width=\linewidth]{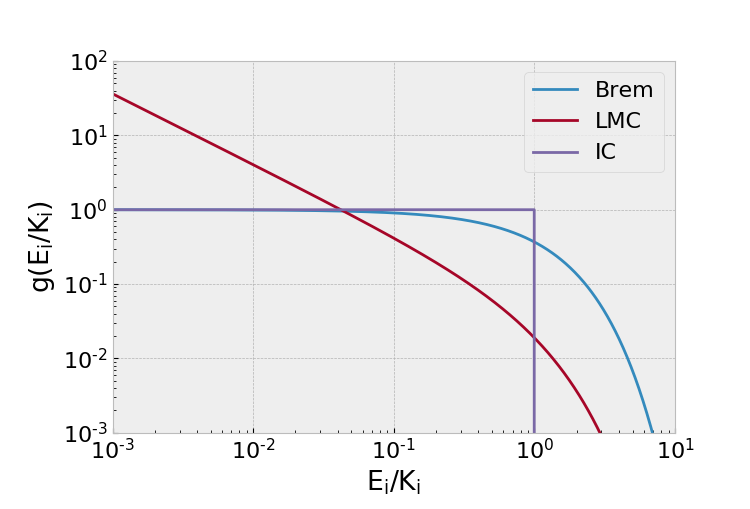}
        \caption{The mathematical parametrization of Bremsstrahlung (\textit{Brem}), laser in a micro-channel (\textit{LMC}), and inverse Compton (\textit{IC}) $\gamma$ sources, defined in table \ref{tab:2.1-ED_definitions} by the functions $g(E_i/K_i)$, are represented against $E_i/K_i$.}
        \label{fig:2.2-photon_EDF}
        \vspace{0.8cm}
    \end{minipage}
    \begin{minipage}[h]{0.45\linewidth}
        \centering
        \includegraphics[width=\linewidth]{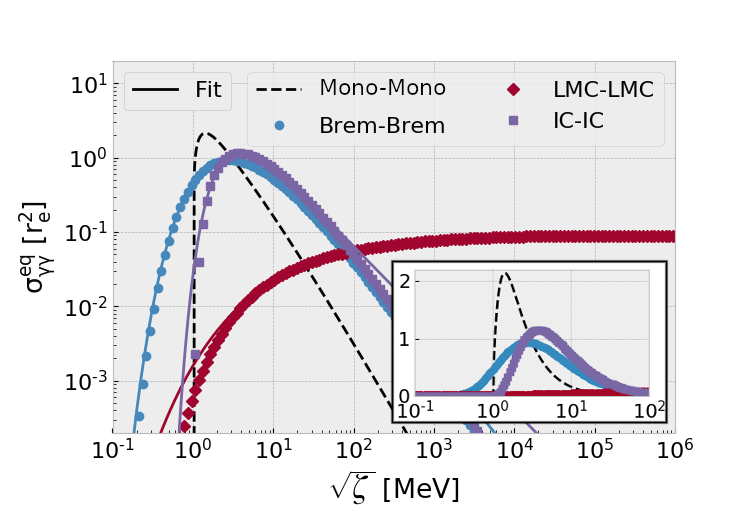}
        \caption{Integrated cross section (in $r_e^2$), calculated from equation (\ref{eq:2.1-ECS_g_g}) with various input energy distribution functions (defined in table \ref{tab:2.1-ED_definitions}), plotted against $\sqrt{\zeta}$ (in MeV). A fit of the form of equation (\ref{eq:2.2-h_definition}) is applied to the numerically calculated values. The LBW cross section, defined in equation (\ref{eq:2.1-LBW_cross_section}) is also plotted in dashed black line.}
        \label{fig:2.2-ECS_results}
    \end{minipage}
\end{figure}

The $g(E_i/K_i)$ functions for these situations, given in table \ref{tab:2.1-ED_definitions}, are also plotted against $E_i/K_i$ in figure \ref{fig:2.2-photon_EDF}. 
The \textit{Brem} and \textit{LMC} energy distribution functions are monotonously decreasing while the \textit{IC} energy distribution function is constant up to a cutoff energy. 

\begin{table}[h]
\centering
\begin{tabular}{ | l | l | l | l | l | }
\hline
    & Type of EDF (\textit{Abbreviation})
                & Mathematical form of $g(E_i/K_i)$
                & Behaviour\\
\hline
i.  & Bremsstrahlung (\textit{Brem})
                & $\exp{(-E_i/K_i)}$
                & Monotonic decrease\\[.3cm]
ii. & Laser + micro-channel (\textit{LMC})
                & $ \dfrac{1}{\Gamma(0.05)} \left(\dfrac{E_i}{K_i}\right)^{-0.95} \exp\left(-\dfrac{E_i}{K_i}\right)$
                & Monotonic decrease\\[.4cm]
iii.& Inverse Compton (\textit{IC})
                & $\left\{
                      \begin{array}{@{}ll@{}}
                        1 & \text{if}\ E_i/K_i < 1 \\
                        \vspace{-2mm}\\
                        0 & \text{otherwise}
                      \end{array}\right.$
                & Constant up to a cutoff\\[.35cm]
\hline
\end{tabular}
\caption{Energy distribution function definitions for the calculation of the integrated cross sections. The mathematical form of $g(E_i/K_i)$, satisfying the conditions given in equations (\ref{eq:2.1-g_definition}) and (\ref{eq:2.1-f_normalization}), is given. $\Gamma$ is the gamma function.}
\label{tab:2.1-ED_definitions}
\end{table}

\subsubsection{Integrated cross section for selected photon sources}
Using these definitions, we can then study the interaction of two Bremsstrahlung sources, two sources produced by the propagation of a laser in a micro-channel, or of two linear inverse Compton sources by inserting the definitions of $g_i$ given in table \ref{tab:2.1-ED_definitions} into equation (\ref{eq:2.1-ECS_g_g}). This integration have been performed numericaly for respectively two \textit{Brem}, two \textit{LMC} and two \textit{IC} $\gamma$ energy distribution functions, with 300 logarithmically spaced values of $\zeta$ between $10^{-2}$ and $10^{12} ~ \rm MeV^2$. 
It is also interesting to note that the cross section for the interaction of two beams with no angular divergence and mono-energetic energy distribution, i.e. described by Dirac delta energy distribution $f_i(E_i, K_i) = \delta(E_i/K_i - 1)$, can be deduced directly from the definition of equation (\ref{eq:2.1-ECS_definition}). Indeed, the integrated cross section for the interaction of such mono-energetic beams (later denoted as \textit{Mono}) reduces to the LBW cross section (defined in equation (\ref{eq:2.1-LBW_cross_section})), where the Mandelstam variable $s$ and our variable $\zeta$ are equivalent. 
These cross sections are plotted against $\sqrt{\zeta}$ in MeV in figure \ref{fig:2.2-ECS_results}, with a zoom on the most interesting region of $\sqrt{\zeta} \in [10^{-1}, 10^2]$ MeV.


When comparing the \textit{Mono}-\textit{Mono} cross section with integrated cross section obtained with wider EDF (such as two \textit{Brem}, \textit{LMC} and \textit{IC} here), we can note that for the collision of beams with wide energy distributions, the integrated cross section shape is wider and has a smaller amplitude.
This means that, all else being equal, a variation in characteristic energy or collision angle will then have a limited effect on the total number of produced pairs.
On the other hand, the decreasing of the amplitude shows that the maximum number of pairs produced is smaller for the collision of two beams with wide energy distributions than it would be for mono-energetic beams, especially for the two \textit{LMC} case. The reduction factor could however be compensated by a corresponding increase in the luminosity (see equation (\ref{eq:2.1-N+_definition})).
To explain these behaviours qualitatively, we can firstly note that in the collision of two mono-energetic beams with no angular divergence, all the binary collisions involve the same collision angle and the same energies of the incoming photons, and so the same value for the variable $s=2 E_1 E_2 (1-\cos \psi_{12})$. The pair production probability (modelised by the LBW cross section, see equation (\ref{eq:2.1-LBW_cross_section})) is then identical in each binary collision, and can be maximized for properly chosen collision angle and photons energies.
In the collision of beams with wide EDF, the energy couple of the photons involved in each binary collision is however not always the same and is asigned to a given probability. It is then impossible to maximize the LBW cross section in each binary collision, and the total pair production probability (modelised by the integrated cross section) is always smaller than the one corresponding to the optimized interaction of two mono-energetic beams.

As we can see from figure \ref{fig:2.2-ECS_results}, all the represented cross section (involving two \textit{Mono}, two \textit{IC}, two \textit{Brem} or two \textit{LMC} energy distributions) have a small value for low $\zeta$ and then increases up to a maximum, to finally decrease as $\zeta$ increase. The cross sections modeling the interaction of two \textit{Mono} or two \textit{IC} energy distributions also seems to exhibit a threshold behaviour near $\sqrt{\zeta}=2 m_e c^2$.

For the interaction of two mono-energetic beams (case \textit{Mono-Mono}), the energy distribution of each beam is represented by a Dirac delta distribution $f_i(E_i, K_i) = \delta(E_i/K_i - 1)$, and the corresponding cross section is strictly equivalent to the LBW cross section (see equation (\ref{eq:2.1-LBW_cross_section})), with $s$ and $\zeta$ being identical. This cross section then exhibit a threshold for $\zeta \geq (2 m_e c^2)^2$, which is a necessary condition caused by the energy conservation. The pair production probability then increases up to $\zeta \approx 2.06 ~ \rm MeV^2$ and finally decreases for larger value of $\zeta$, identically as the LBW cross section do.

\begin{figure}[t]
    \centering
    \includegraphics[width=\linewidth]{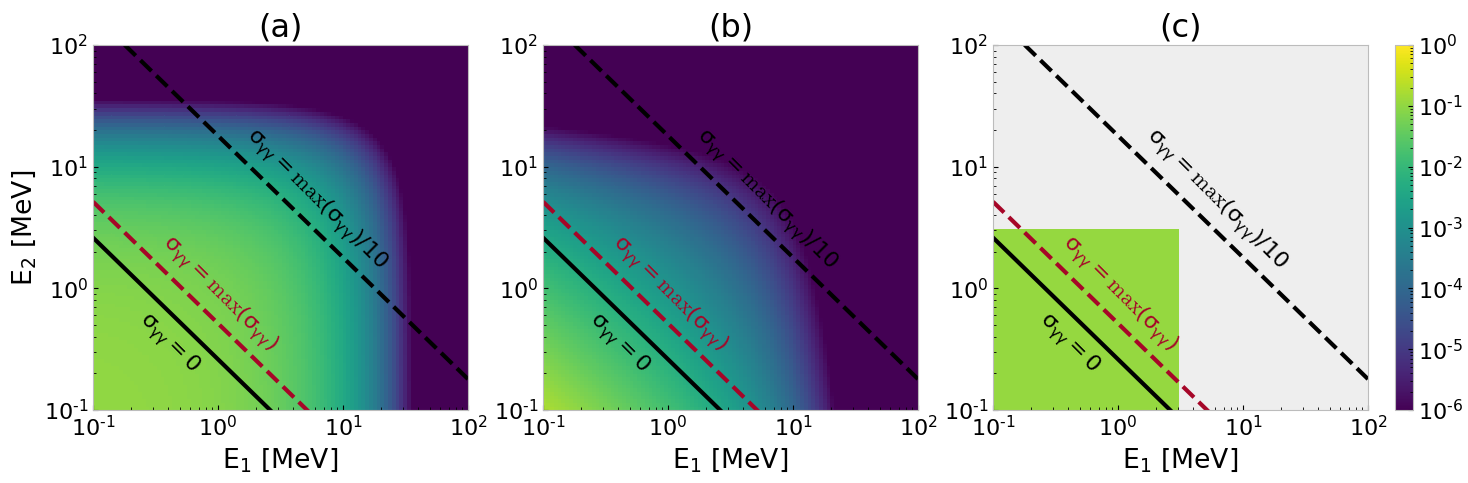}
    \caption{Probability density to get a given $\gamma$ energy couple between $(E_1, E_2)$ and $(E_1 + dE_1, E_2 + dE_2)$ (product $f_1 \times f_2$) for (a) two \textit{Brem}, (b) two \textit{LMC} and (c) two \textit{IC} energy distribution functions, with $K_1=K_2=3$ MeV, in term of photons' energies $E_1, E_2$ (in MeV). The color scale is lower bounded at $10^{-6} ~ \rm MeV^{-2}$ for helping the visualisation. The continuous black line indicates the pair production threshold ($s = (2 m_e c^2)^2$), the dashed red line the maximum cross section ($s \approx 2.06 ~ \rm MeV^2$) and the dashed black line the position where the cross section is only $1/10$ of its maximum ($s \approx 71.8 ~ \rm MeV^2$) assuming $\psi_{12}=180^\circ$. For smaller angles, identical values of $s$ need higher photons energies so these lines move toward the upper right corner of the graphs. The corresponding value of $\zeta$ is here $36 ~ \rm MeV^2$ for the three graphs.}
    \label{fig:2.2-examples_wide_EDF}
\end{figure}

This threshold behaviour is also present in the cross section representing the interaction of two \textit{IC} energy distributions. Indeed, the probability density of such energy distribution is constant up to a high energy cutoff, noted $K_1$ and $K_2$ respectively for beams $1$ and $2$ (see table \ref{tab:2.1-ED_definitions}). For any value of $K_1$, $K_2$ and $\psi_{12}$, the maximum value of the Mandelstam variable $s=2E_1 E_2 (1-\cos \psi_{12})$ achievable in any binary collision is thus limited to $2 K_1 K_2 (1-\cos\psi_{12})$. This expression being literally the definition of our variable $\zeta$, the threshold behaviour of this integrated cross section in figure \ref{fig:2.2-ECS_results} can be explained by the fact that for $\zeta < (2 m_e c^2)^2$ no binary collision could satisfy the pair production threshold $s\geq (2 m_e c^2)^2$. 
For higher values of $\zeta$, the increase of the integrated cross section can be explained by an increasing proportion of the binary collisions satisfying the two-photon pair production threshold (given by $s\geq (2 m_e c^2)^2$). An example of such behaviour is illustrated in figure \ref{fig:2.2-examples_wide_EDF}c, reprensenting the probability density for a binary collision to involve photons energies between $(E_1, E_2)$ and $(E_1 + dE_1, E_2 + dE_2)$ (given by the value of the product $f_1 \times f_2$), for $K_1=K_2=3$ MeV. The solid black line in this figure indicates the pair production threshold, the dashed red line indicates the position of the maximum cross section, and the dashed black line the position where the value of the cross section is $1/10$ of its maximum, assuming a collision angle $\psi_{12} = 180^\circ$. 
The integrated cross section can then be understood as the product of this function with the LBW cross section, integrated over energies for a given angle, such as stated in equation (\ref{eq:2.1-ECS_with_dependencies}). As we can see from this figure, not all the photon binary collisions can contribute to the pair production (only the ones situated higher than the two-photons pair production threshold can). 
A situation corresponding to $\zeta < (2 m_e c^2)^2$ could also have been illustrated with the same kind of figure, and would imply that all the region where the probability is non-zero would be situated under the pair production threshold, at the lower left corner of the plot.
On the opposite, for larger values of $K_1$ and $K_2$ (i.e. for large $\zeta$) the integrated cross section decreases with increasing $\zeta$. Indeed, even if the proportion of the binary collision satisfying the pair production condition $s\geq (2 m_e c^2)^2$ can be important, most of them involves large values of $s$, which are not very efficient to produce pairs (see figure \ref{fig:2.1-LBW_cross_section}).

The definitions of the \textit{Brem} and \textit{LMC} energy distributions (see table \ref{tab:2.1-ED_definitions}), do not involve any high energy cutoff this time, and these functions are  continuous and monotonously decreasing (see figure \ref{fig:2.2-photon_EDF}). We then note no threshold behaviour in the integrated cross sections plotted in the figure \ref{fig:2.2-ECS_results}, where the characteristic energies $K_1$, $K_2$ involved in the definition of $\zeta$ refer this time to the slope (or "effective temperature") of such energy distributions (see table \ref{tab:2.1-ED_definitions}). The value of the integrated cross section is however small for small values of $\zeta$, because small values of $K_1$ and $K_2$ tends to favor the lowest energy photons, which are less likely to satisfy the pair production threshold $s \geq (2 m_e c^2)^2$.
At fixed collision angle, for increasing values of $K_1$ and $K_2$ the proportion of the binary collisions higher than the pair production threshold $s\geq (2 m_e c^2)^2$ increases. This effect leads to an increase of the integrated cross section with $\zeta$ for all the studied cases with wide energy distributions. An example is also represented in figures \ref{fig:2.2-examples_wide_EDF}a and \ref{fig:2.2-examples_wide_EDF}b respectively for two \textit{Brem} and two \textit{LMC} energy distributions, and for $K_1=K_2=3 ~ \rm MeV$.
For higher values of $\zeta$, the integrated cross sections decreases because most of the binary collisions are situated in a zone which has a low pair production probability, such as indicated by the dashed black lines in figures \ref{fig:2.2-examples_wide_EDF}.

\subsubsection{Implications for two photons pair production experiments}
For these configurations, the integrated cross section thus goes through a maximum value, denoted $\zeta'$. In order to maximize the number of pairs produced in the interaction of these photon beams, one should maximize the product of the luminosity by the integrated cross section (see equation (\ref{eq:2.1-N+_definition})). For given $\gamma$ production mechanism, the total number, the spatio-temporal characteristics and the energy distribution of the $\gamma$ source can however vary simultaneously with experimental parameters, such as e.g. the laser intensity. The maximum number of pairs can then possibly be obtained for sub-optimum value of integrated cross section, if a sufficient increase in luminosity compensates the decrease in cross section. Nevertheless, the optimum characteristic energies obtained from equation (\ref{eq:2.1-optimum_S}) remain useful estimates for the design of such pair production experiments.

In the following examples, we will discuss the values of the characteristic energies that maximizes the integrated cross section, when considering two counter-propagating ($\psi_{12}=180^\circ$) identical photon beams ($K_1=K_2$). With these conditions and the equation (\ref{eq:2.1-optimum_S}), we then have: 

\begin{enumerate}[label={\Roman*. }]
    \item For the interaction of two Bremsstrahlung sources, our estimates from equation (\ref{eq:2.1-optimum_S}) and the value of $\zeta'$ taken from table \ref{tab:2.2-fit_results} gives optimum characteristic energies of about $1.4$ MeV. In high intensity laser-plasma experiments, this kind of characteristic energies have already been experimentally reported since the late 1990's \cite{norreys_1999}.
    
    For a laser of about 80 J of energy interacting with a millimeter-size tantalum solid target, the number of photons produced by Bremsstrahlung is about $6 \times 10^{11}$, in a half angle of divergence of about $10^\circ$, with an initial source diameter of about $80$ µm \cite{palaniyappan_2019}. After a propagation distance of $250$ µm, the typical source surface is thus about $\pi \times (84 ~ \textrm{[µm]})^2$. If we consider the collision of two such photon sources at a distance of 500 µm (each source is located at 250 µm of the collision point) with a collision angle of $180^\circ$, the typical luminosity of such configurations is about $(6 \times 10^{11})^2/(\pi (84 ~ \textrm{[µm]})^2) \sim 2 \times 10^{27} \rm cm^{-2}$, and for an integrated cross section of $r_e^2$ the corresponding number of LBW pairs is about $200$ per shot. This number is then significant, even if it is about 2 orders of magnitudes lower than the prediction given in reference \cite{ribeyre_2016} with similar parameters. 
    From these considerations, using Bremsstrahlung sources can then be promising to observe LBW pair production.
    
    However, in this kind of sources the angular divergence of the $\gamma$ photons of about tens of degrees \cite{henderson_2014} is not negligible, and the $\gamma$ source is not homogeneous \cite{norreys_1999}, invalidating some of our previous assumptions. Our model should then not be seen as an accurate description of the LBW pair production by the collision of two laser-based Bremsstrahlung $\gamma$ sources, but more as an helpful tool to quickly estimate the most suitable $\gamma$ sources between in a wide range of possibilities. The propagation of $e^-$ or $\gamma$ photons in a high $Z$ material can also create a large number of background $e^+$ via the Bethe-Heitler ($\gamma Z \to Z e^- e^+$) or Coulombian Trident ($e^- Z \to e^- Z e^- e^+$) processes \cite{ruffini_2010}, potentially interfering with the detection of the LBW process.
    
    The energy distribution of Bremsstrahlung $\gamma$ sources can also be possibly modeled by other EDF satisfying both equations (\ref{eq:2.1-g_definition}) and (\ref{eq:2.1-f_normalization}), or using a sum of exponentials such as used e.g. in reference \cite{zulick_2013}. A brief discussion on the usage of a sum of functions to estimate total pair production will be given at the very end of this section. 

    \item Concerning the integrated cross section for the two \textit{LMC} configuration, equation (\ref{eq:2.1-optimum_S}) gives the optimum characteristic energies for two identical counter-propagating beams of about $4.9 \times 10^4$ MeV. This value is well beyond of the values obtained from the fit of the data of reference \cite{wang_2020}, which is about few MeV (for the low energy part) or tens of MeV (for the high energy tail). However, as can be seen from figure \ref{fig:2.2-ECS_results}, the value of the integrated cross section in this case is almost constant at high value of $\zeta$, and is about $1/2$ of its maximum value for $\zeta \approx 43 ~ \rm MeV^2$ where the values of the optimum characteristic energies are more reasonable (about $3.3$ MeV, similar to what was found for the low energy part of Wang's data). 
    
    Because of the smallness of the integrated cross section, this configuration can then firstly seem to be not very adapted to the LBW pair production, but the high luminosity that could be produced by the interaction of such $\gamma$ sources could eventually overcome this difficulty. 
    
    As an example, the typical number of $\gamma$ photons with energy $>10$ keV produced by 2 PW laser pulse (75 J of energy) is about $2 \times 10^{12}$, within a typical diameter of about the laser spot size (3 µm) and with a divergence half-angle that have been considered to be about $5^\circ$ for the estimates made in this reference \cite{wang_2020}. If we suppose a collision distance of $500$ µm (each source is located at 250 µm to the collision point), a collision angle of $180^\circ$ and a value of the integrated cross section of about $0.1 r_e^2$, the luminosity of such collision is about $2 \times 10^{29} \rm cm^{-2}$ and the number of produced LBW pairs is about $2000$ per shot. This number is then very similar to what have been reported in this paper with the same collision distance and a collision angle of $90^\circ$.
    
    Thus, despite a smaller integrated cross section, the number of pairs produced by the interaction of such photon sources is here one order of magnitude greater than the one predicted for Bremsstrahlung sources, with similar laser energies (see previous point) and with an expected much lower level of background positrons produced by other process \cite{ribeyre_2016}. This kind of source could then also be a credible candidate for such photon collision experiments. 
    
    These estimates however neglects the internal structure of the photon beams, which is also very important for these kind of $\gamma$ sources. Our parametrization of the Wang's data by the function defined in table \ref{tab:2.1-ED_definitions} is also not very accurate in the higher energy part of the energy distribution (higher than about 10 MeV), and could be improved e.g. by using the sum of two such functions (one for the low energy part and one for the higher energies).
    
    \item For the collision of two \textit{IC} energy distributions, the optimum characteristic energies given by our model for two identical counter-propagating beams is about $2.0$ MeV. This kind of photon energies have already been reported in the litterature for a single photon beam \cite{albert_2016}.
    
    In their paper, \textit{Drebot et al., 2017} \cite{drebot_2017} consider two identical counter-propagating beams with energies up to $K_1 = K_2 \approx 1.2$ MeV, which is sub-optimal in our modelisation ($\zeta \approx 5.8 ~ \rm MeV^2$). The decrease on integrated cross section is however only limited to a factor $\sim 1.2$.
    To get the optimum integrated cross section assuming these conditions and the same laser as the authors ($1 ~ \rm \mu m$ wavelength), the required initial electron beam energy should be about $325$ MeV, instead of the $260$ MeV considered in their paper.

    By using the parameters of the photons source from this paper, the typical luminosity of the collision of such beams is about $2 \times 10^{23} \rm cm^{-2}$ for a collision distance of 8 mm (we supposed a typical source diameter of 10 µm, a divergence half-angle of $0.3^\circ$ and a number of photons of $2 \times 10^9$ taken from another reference with similar parameters \cite{micieli_2016}). For an integrated cross section of about $r_e^2$, the total number of produced pairs is then predicted to be about $2 \times 10^{-2}$ per shot, which is 2 order of magnitudes higher than the number given in their paper obtained by the mean of numerical simulations.
    
    As these authors discussed in their paper, the photons with higher energies are however the most collimated along the propagation axis, invalidating the assumption we made on equation (\ref{eq:2.1-homogeneous_beam}). We also considered a perfectly constant energy distribution up to a cutoff energy, which is only a rough estimate of the photon energy distribution given in this paper.
    Our results should then again be seen as very rough estimates more than as accurate description of what may happen in the interaction of two of such inverse Compton $\gamma$ beams.
    
\end{enumerate}

The obtained results for $\sigma_{\gamma\gamma}^{int}$ were fitted using the function $h(\zeta)$ defined by:

\begin{equation}
    h(\zeta) = r_e^2 \times (a / \zeta^n) \times \exp\left(-b / \zeta^m \right) ~ \rm ,
    \label{eq:2.2-h_definition}
\end{equation} 

where $a$, $b$, $n$ and $m$ are positive real numbers. As shown on figure \ref{fig:2.2-ECS_results}, the fits agree reasonably well for values of integrated cross section above $10^{-3} ~ r_e^2$, and are very accurate nearby the maximum of the integrated cross sections. The fit parameters obtained for $\zeta$ in $\rm MeV^2$ are given on table \ref{tab:2.2-fit_results}.

\begin{table}[htbp]
\centering
\begin{tabular}{ | l | l || l | l | l | l || l | l |}
\hline
    & Beams EDF   
    & $a$ [$\rm MeV^{2n}$]      & $b$ [$\rm MeV^{2m}$]
    & $n$       & $m$       
    & $\zeta'$ [$\rm MeV^2$]  & $\sigma_{\gamma\gamma}^{int}(\zeta')$ [$r_e^2$]
    \\
\hline
I.  & \textit{Brem} - \textit{Brem}
    & $23.3$    & $3.94$        
    & $0.674$   & $0.374$   
    & $8.00$    & $0.94$    
    \\
II. & \textit{LMC} - \textit{LMC}
    & $9.82 \times 10^{-2}$ & $4.13$         
    & $3.32 \times 10^{-3}$ & $0.221$   
    & $9.70 \times 10^9$    & $8.93 \times 10^{-2}$
    \\
III.& \textit{IC} - \textit{IC}
    & $10.5$    & $6.01$        
    & $0.552$   & $0.798$   
    & $15.3$    & $1.15$
    \\
\hline
\end{tabular}
\caption{Values of the parameters of the fit function defined in equation (\ref{eq:2.2-h_definition}) for the specified integrated cross section.}
\label{tab:2.2-fit_results}
\end{table}

Using these parameters for the function $h$, one can then quickly and accurately estimate $\sigma_{\gamma\gamma}^{int}$ in a wide range of configurations without performing the integration of equation (\ref{eq:2.1-ECS_g_g}), and estimate the total number of produced pairs by replacing $\sigma_{\gamma\gamma}^{int}$ by $h$ in equation (\ref{eq:2.1-N+_definition}).

Especially, it is interesting to note that this fitting function could be used to model energy distribution functions that does not directly satisfies the condition given by equation (\ref{eq:2.1-g_definition}), but are composed of a sum of functions that do so. 
As an example, let us consider that the energy distribution of the beam $1$ can be described by a function satisfying both equations (\ref{eq:2.1-g_definition}) and (\ref{eq:2.1-f_normalization}), and that the energy distribution of the beam $2$ can be described by a two-component EDF such as:
\begin{equation}
    \tilde{f_2}(E_2, K_{2a}, K_{2b}) = p \times f_{2,a}(E_2,K_{2a}) + (1-p) \times f_{2b}(E_2,K_{2b}) ~ \rm ,
\end{equation}
where $K_{2a}$ and $K_{2b}$ are two characteristic energies, $f_{2a}$ and $f_{2b}$ are functions that satisfies both equations (\ref{eq:2.1-g_definition}) and (\ref{eq:2.1-f_normalization}), and $p \in[0,1]$ defines the relative weight of each of the two component in the energy distribution of the beams. An example of such two component EDF $\tilde{f_i}$ could be a sum of two exponentials (as defined in table \ref{tab:2.1-ED_definitions}) modeling respectively the low energy part and the high energy part of Bremmstrahlung energy distributions (see e.g. reference \cite{zulick_2013}). 
By inserting this composite energy distribution function into the basic definition of $\sigma_{\gamma\gamma}^{int}$ given in equation (\ref{eq:2.1-ECS_definition}), it is straightforward to show that the integrated cross section can be expressed as a sum of two terms, whom are in the form of equation (\ref{eq:2.1-ECS_with_dependencies}). Each of these term can then be approximated by the fitting function $h$, and the resulting integrated cross section can then be estimated as:
\begin{equation}
    \sigma_{\gamma\gamma}^{int} \approx p \times h_{a}(\zeta_a) + (1-p) \times h_{b}(\zeta_b) ~ \rm ,
\end{equation}
where $h_a$ and $h_b$ are the fitting functions corresponding to the interaction of one EDF ($f_1$ in our example) with the first component of the composite EDF (here $f_{2a}$) or the second component of the composite EDF (here $f_{2b}$) respectively, and where $\zeta_a=2 K_1 K_{2a} (1-\cos\psi_{12})$ and $\zeta_b=2 K_1 K_{2b} (1-\cos\psi_{12})$. 
The same kind of reasoning could be eventually generalized to treat more complex situations from these basic results. 


In this section, we have seen a general, semi-analytical method to estimate the total number of $e^-e^+$ pairs produced by the LBW process in the collision of two directional beams of real $\gamma$ photons with simple energy distributions and no angular divergence. 
We have then parametrized Bremsstrahlung and multiphoton inverse Compton scattering laser-based $\gamma$ sources, and inverse Compton $\gamma$ source generated by the coupling of two laser with a lepton collider, with such simple energy distribution functions.
We have determined optimized energy distribution parameters to optimize the total pair production by the LBW process with such photon sources. These theoretical optimums have shown that, from an energetic point of view, optimum Bremsstrahlung sources are already reachable in typical high intensity laser-plasma experiments since the late 1990's. The photon energy distributions of the proposition of \textit{Drebot et al., 2017} are close to the optimum but slightly sub-optimal in our model. For the proposition of \textit{Wang et al., 2020} involving the collision of two photon sources created by a laser propagating in a micro-channel, the photons energy distribution is, at a first glance, less interesting for LBW pair production. However, the high collision luminosity of such configuration could significantly overcome the less interesting energy distributions features. 
Fitting functions reproducing the value of the integrated cross section, taking into account all the $\gamma$ energy couplings, were also given for the three studied cases. This model could also be adapted to other energy distribution functions.

\section{Conclusion}

Despite its fundamental importance in quantum electrodynamics and astrophysics, the process of pair production by two real photon collision has yet still to be detected in the laboratory. However, various propositions emerged to study this process in laboratory in the last few years, based upon various photon production mechanisms, and thus various photons energy distributions. These energy distribution effects have been taken into account by the mean of analytical model only in two of these proposition concerning the produced pair number.

The aim of the present paper was to give a general and original, semi-analytical method to estimate the influence of the incoming photon energy distribution in the produced pair number and kinematics, given the photon beam have simple energy distributions and no angular divergence.

In some conditions, the total number of pairs produced in the collision of two such photon beams can simply be expressed as a product of a geometric luminosity with a so-called integrated cross section, taking into account all the energy couplings of the photon beam collision. For simple photon energy distributions, this integrated cross section can be reduced to a one variable function, allowing to optimise energy and angular parameters of the collision easily.

Our results shows that inverse Comptonized laser photon beams in a lepton collider, and laser-based Bremsstrahlung sources are, only from an energetic point of view, very efficient to produce pairs, and are also already reachable with today's facilities. The photons produced by the propagation of a PW-class laser in a micro-channel seems at first glance less interesting for the application of two photon pair production, but the high collision luminosity of such setup can eventually overcome the less efficient energy couplings. Our assumptions are however very restrictive, and these results should be seen more as helpful order of magnitudes estimates than as quantitative results of what would happen in such experiments. A fitting function reproducing our numerical results is also given. 

These results could be immediately adapted to other photon energy distributions, providing their formulation satisfies few simple criteria, allowing to study other laboratory situations or astrophysical situation like for example super-massive black hole environment in active galactic nuclei. This original method could also be adapted to photo-production of heavier leptons (such as muons) simply by rescaling the energies and cross sections, or even eventually to other processes involving photons or other kind of colliding particles.

\section*{Acknowledgments}
This research was supported by the French National Research Agency (Grant No. ANR-17-CE30-0033-01) TULIMA Project and by the NSF (Grants No. 1632777 and No. 1821944) and AFOSR (Grant No. FA9550-17-1-0382).
We acknowledge T Wang for the data he provided for the analysis of the \textit{LMC} case, and for his fruitful advices.
\printbibliography
\end{document}